\documentclass[prb,floatfix,superscriptaddress,amsmath,amssymb,twocolumn,nobalancelastpage]{revtex4}

\usepackage[utf8]{inputenc}
\usepackage{graphicx}
\usepackage{dcolumn}
\usepackage{bm}
\usepackage{datetime}
\usepackage{amsmath} 
\usepackage{float}
\usepackage{color}
\usepackage{dsfont}
\usepackage{lipsum}
\usepackage[normalem]{ulem}
\usepackage{hyperref}
\begin{document}

\title{On the neural network flow of spin configurations}

\author{Santiago Acevedo}
\email{santiagoacevedo@fisica.unlp.edu.ar}
\affiliation{IFLP - CONICET, Departamento de F\'isica, Universidad Nacional de La Plata,
C.C.\ 67, 1900 La Plata, Argentina.}
\affiliation{CCT CONICET La Plata, Consejo Nacional de Investigaciones Cient\'\i{}ficas y T\'ecnicas, Argentina}

\author{Carlos A.~Lamas}
\affiliation{IFLP - CONICET, Departamento de F\'isica, Universidad Nacional de La Plata,
C.C.\ 67, 1900 La Plata, Argentina.}
\affiliation{CCT CONICET La Plata, Consejo Nacional de Investigaciones Cient\'\i{}ficas y T\'ecnicas, Argentina}
\affiliation{Departamento de F\'\i{}sica, Facultad de Ciencias Exactas, Universidad Naiconal de La Plata, Argentina}

\author{Alejo Costa Duran}

\affiliation{CCT CONICET La Plata, Consejo Nacional de Investigaciones Cient\'\i{}ficas y T\'ecnicas, Argentina}
\affiliation{Instituto de F\'isica de L\'\i{}quidos y Sistemas Biol\'ogicos (IFLySiB) --- Universidad Nacional de La Plata and CONICET, Calle 59 n.\ 789, B1900BTE La Plata, Argentina.}

\author{Mauricio B.~Sturla}

\affiliation{CCT CONICET La Plata, Consejo Nacional de Investigaciones Cient\'\i{}ficas y T\'ecnicas, Argentina}
\affiliation{Instituto de F\'isica de L\'\i{}quidos y Sistemas Biol\'ogicos (IFLySiB) --- Universidad Nacional de La Plata and CONICET, Calle 59 n.\ 789, B1900BTE La Plata, Argentina.}

\author{Tom\'as S.~Grigera}

\affiliation{CCT CONICET La Plata, Consejo Nacional de Investigaciones Cient\'\i{}ficas y T\'ecnicas, Argentina}
\affiliation{Departamento de F\'\i{}sica, Facultad de Ciencias Exactas, Universidad Naiconal de La Plata, Argentina}
\affiliation{Instituto de F\'isica de L\'\i{}quidos y Sistemas Biol\'ogicos (IFLySiB) --- Universidad Nacional de La Plata and CONICET, Calle 59 n.\ 789, B1900BTE La Plata, Argentina.}
\affiliation{Istituto dei Sistemi Complessi, Consiglio Nazionale delle Ricerche, via dei Taurini 19, 00185 Rome, Italy}


\begin{abstract}

We study the so-called neural network flow of spin configurations in the 2-$d$ Ising ferromagnet.  This flow is generated by successive reconstructions of spin configurations, obtained by an artificial neural network like a restricted Boltzmann machine or an autoencoder.  It was reported recently that this flow may have a fixed point at the critical temperature of the system, and even allow the computation of critical exponents.  Here we focus on the flow produced by a fully-connected autoencoder, and we refute the claim that this flow converges to the critical point of the system by directly measuring physical observables, and showing that the flow strongly depends on the network hyperparameters.  We explore  the network metric, the reconstruction error, and we relate it to the so called intrinsic dimension of data, to shed light on the origin and properties of the flow.

\end{abstract}

\maketitle

\section{Introduction}

Machine learning (ML) techniques have in recent years found application in many problems in physics, and are being increasingly adopted by the physics community. The most common application is the study of phase transitions \cite{carrasquilla2017machine, van2017learning, CORTE2021, Ponte2017, Wang2016, Wang2017, IntrinsicDimension2021}, but ML has also been used to study problems such as the prediction of crystal structures \cite{fischer2006predicting, Crystal2003}, the processing of neutron scattering data \cite{samarakoon2020}, the speed-up of Monte Carlo (MC) simulations \cite{RBM-MonteCarlo}, the renormalization  group (RG) transformation \cite{koch2018mutual,RG-ML}, quantum state tomography \cite{torlai2018neural}, and many-body quantum states encoding \cite{carleo2017}, 
among many others\cite{Review-ML-physics}.  The relationship between ML and RG, and the possible implementation of RG schemes through neural networks (NNs) is particularly intriguing, as it brings together what is arguably the most important single concept of statistical physics of the second half of the XXth century with a field of artificial intelligence that has experienced spectacular growth in the last decade.

\begin{figure}[t]
    \centering
    \includegraphics[width= \linewidth]{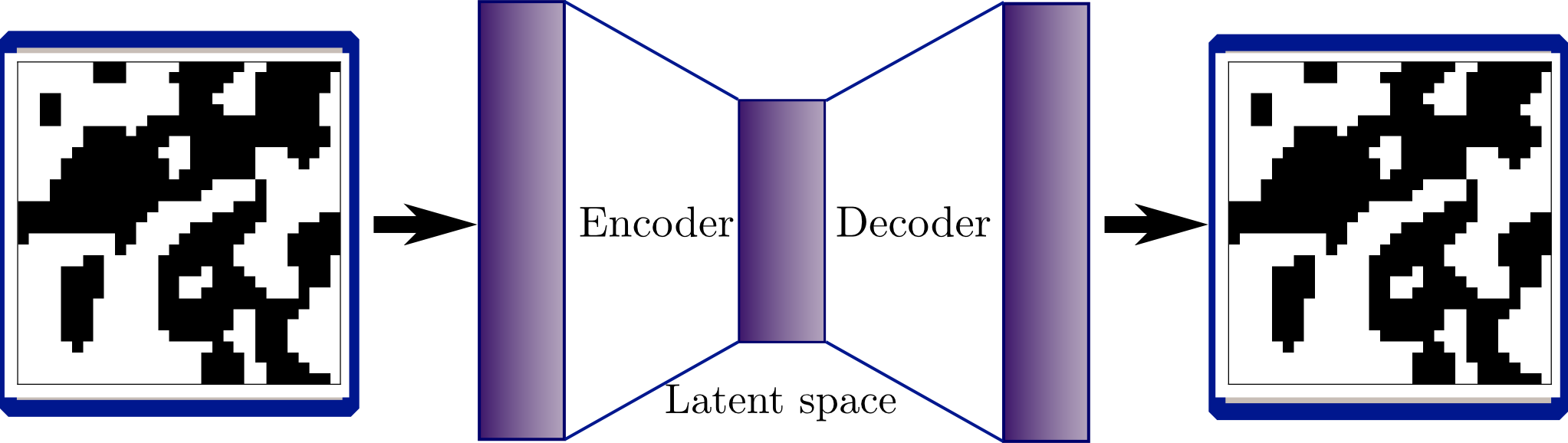}
    \caption{Diagram of an AE. To the left, the input layer. At the center, the latent space, where data is compressed. At the right, the output layer. }
    \label{fig:AE}
\end{figure}

\begin{figure*}[t]
    \centering
    \includegraphics[width= \linewidth]{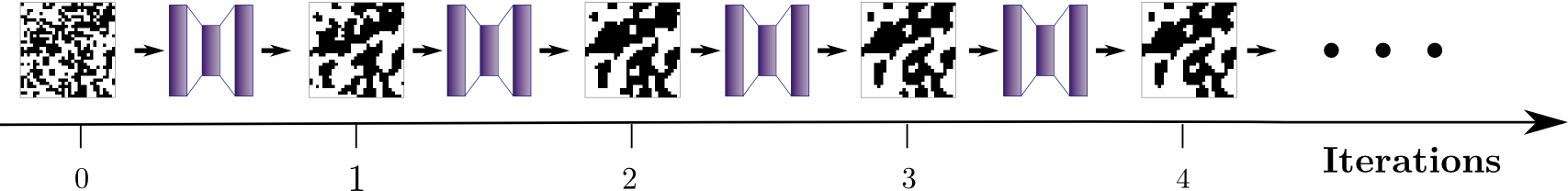}
    \caption{Diagram of the AE flow, generated by consecutive reconstructions of spin configurations made by a trained AE. }
    \label{fig:flujo}
\end{figure*}

The recently introduced notion of Neural Network flow \cite{RBMFlow0, RBMFlow} provides in principle a possible realization of the coarse-graining transformation, which is the first step of the RG.  The idea is that a NN trained to reproduce images, such as an autoencoder (AE), can be fed an image to be reproduced, and this reproduction can be fed back into the NN, thus generating a sequence (See Fig. \ref{fig:AE}, \ref{fig:flujo} and Sec. \ref{sec:auto-encoder}).  Since the reproduced image contains errors, the procedure generates different images at each iteration, until eventually reaching a fixed point, i.e.\ an image that can be exactly reproduced by the NN.  If the NN is an AE with a hidden layer smaller than the input layer, this dimensional reduction forces the NN to do a sort of coarse-graining of the image.  The result is a flow of coarse-graining-like transformation that is similar to the RG flow, albeit with the lack of the (crucial) rescaling RG step.

This flow was studied in Refs.~\onlinecite{RBMFlow0}, \onlinecite{RBMFlow}, and \onlinecite{giataganas2021neural}, and it was found to be related to the actual RG flow of the physical system under consideration.  In \onlinecite{RBMFlow0} and \onlinecite{RBMFlow}, the NN flow was implemented on a Restricted Boltzmann Machine (RBM) network. 
The network was trained with Monte Carlo configurations of an Ising spin system at various temperatures and magnetic fields. 
%

The flows obtained by feeding back to the network reproduced images were found to have fixed points, which appear to correspond to points in the phase diagram where the Ising models present specific heat maxima. In this manner, the RBM flow would spontaneously identify the phase transition of the Ising model.
In Ref.~\onlinecite{giataganas2021neural} a similar flow was obtained using both autoencoders and variational autoencoders (VAEs). The broad finding of these papers is that RBMs and AEs can be trained and set up to obtain a transformation that acts like an ``anti-RG'', producing a flow that takes images to the critical point.

Here we revisit the NN-flow of the Ising model with a standard AE.  We show that, although certain features of the NN-flow fixed point can be associated with critical configurations of the system, others do not.  We find that NN-flow has non trivial dependencies on the reconstruction error, on the intrinsic dimension of the system\cite{IntrinsicDimension2021}, and on the way in which the NN is trained.  Most importantly, there are a multitude of fixed points in image space, and most of them cannot be said to be related to the unstable fixed point of the RG (i.e.\ the critical point).

\section{Model and neural network}

\subsection{The Ising model}

We test the NN-flow ideas on the very well known Ising ferromagnet in $d=2$ dimensions, with periodic boundary conditions, first-neighbor interactions and no external field.  It is defined by the energy
\begin{equation}
 \mathcal{H} = - J \sum_{\langle i,j \rangle} \sigma_i \sigma_j,
 \label{eq:H}
\end{equation}
where ${\langle i,j \rangle}$ stands for nearest neighbor pairs on a square lattice, $J$ is the spin-spin coupling constant (here set to one), and $\sigma=\pm 1$.  The order parameter is the magnetization per spin,
\begin{equation}
 m= \frac{1}{N} \sum_{i} \sigma_{i},
 \label{eq:m}
\end{equation}
where $N$ is the number of sites. This model presents a continuous transition between the paramagnetic phase and the ordered phase, which in the thermodynamic limit is known to be at $T_c=2/(\ln{(1+\sqrt{2})}) \approx 2.269$.
For this work, we have generated a set of equilibrium configurations on a $30\times30$ square lattice with standard single-spin flip Metropolis Monte Carlo \cite{newman1999monte}.  We have run 400 independent simulations, starting from the disordered phase (initial temperature $T_0=4.5$).  Each simulation was run at a fixed temperature belonging to a set of 200 evenly-spaced temperatures in the range $[0.02,T_0]$ until equilibrium was reached, and the final spin configuration was saved.  Thus we generated a set  of $80000$ configurations, or images.  We used $60000$ configurations as training data, of which 10\% was taken as validation data, to monitor training and choosing network hyperparameters.  The remaining $20000$ configurations ($100$ for each temperature) constitute our test set.

\subsection{The autoencoder}
\label{sec:auto-encoder}

AEs\cite{Goodfellow-et-al-2016} are neural networks whose aim is to make an approximate copy of the given input, $\boldsymbol{x}$. They consist of two parts, encoder and decoder (see Fig. \ref{fig:AE}).  The former generates a representation $\boldsymbol{z}$ of each data point $\boldsymbol{x}$ in the so called latent space, which typically has a lower dimension that the original space in which data is embedded. The latter takes the representation $\boldsymbol{z}$ and reconstructs an approximate copy 
$\bar{\boldsymbol{x}}$ of the encoder input, $ \boldsymbol{x}$.
The AE trainable parameters are learned minimizing the mean square error (mse) between the input and the output for all the elements in the training set, $X$:

\begin{equation}
 mse(X)= \frac{1}{|X| N} \sum_{\boldsymbol{x}\in X} |\boldsymbol{x} - \bar{\boldsymbol{x}}|^2,
 \label{eq:mse}
\end{equation}
where $|X|$ is the number of elements in $X$.  Unlike RBMs, AEs are deterministic functions which map $\boldsymbol{x}$ to $\bar{\boldsymbol{x}}$, and do not learn the underlying probability distribution of the data.  This task can be addressed with VAEs \cite{chollet2021deep,Goodfellow-et-al-2016}, but we do not consider those here.




In this work we use fully connected AEs with a single hidden layer of $N_l$ units, as they are the simplest neural network reported to be capable of generating a flow of configurations towards the critical point \cite{giataganas2021neural}.  The training was done with a fixed number of 50 epochs, batch size of 512, and a learning rate equal to  $10^{-3}$. The trainable parameters are learned minimizing the cost function \eqref{eq:mse} through gradient descent using back propagation.
All neural network calculations are performed with Tensorflow \cite{tensorflow_developers_2021_5799851}.

The input is a vector of 900 components that only take the (normalized) values $0,1$, but the components of the reconstructed copy $\bar{\boldsymbol{x}}$ are real numbers between 0 and 1. 
Therefore, before feeding back the image to the AE or otherwise processing the output, we round each component to 0 or 1. 
As a consequence, when computing the reconstruction error (RE) the rounding procedure implies that every wrong pixel contributes $1/N$ to the mse of a single configuation, being $N$ the number of sites. 
This allows to discretize the RE, and allows the possibility to have perfect reconstructions, where the RE is exactly zero.

\section{Results}

\subsection{AE-flow}
\label{sec:Results-AE-flow}

To study the properties of the NN flow we choose to monitor the energy and magnetization of each configuration using~\eqref{eq:H} and~\eqref{eq:m}.  This is in contrast to Ref.~\onlinecite{giataganas2021neural}, where the authors used the temperature, determined through a different neural network, previously trained to measure the temperature of a configuration.  We prefer to stick to energy and magnetization, which are clearly defined physical variables, rather than resort to an unnecessary black box.  As we shall see, the present approach shows that the NN flow cannot be described by a temperature.

\begin{figure}
    \centering
    \includegraphics[width=\linewidth]{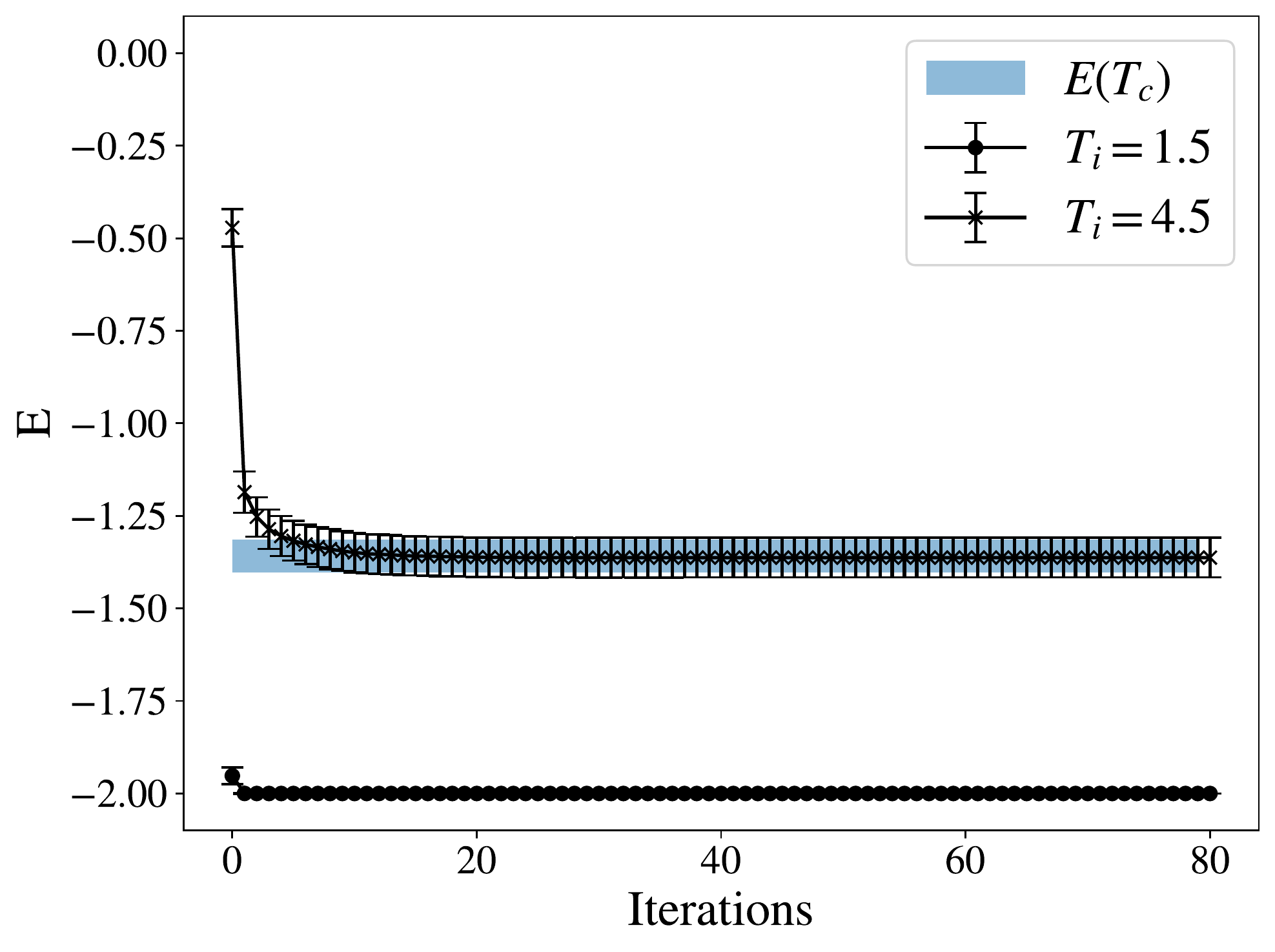}
    \caption{Energy evolution in the NN flow.  We plot the average energy per spin vs.\ the number of iterations, for $100$ independent initial configurations, where error bars correspond to the standard deviation.
    Two different initial temperatures $T_i=4.5$ and $T_i=1.5$ are taken to show two different fixed points. The colored shadow denotes the range of energies within dispersion that corresponds to the critical point of the system.}
    \label{fig:E-flow-128}
\end{figure}

Figure \ref{fig:E-flow-128} shows the energy flow starting from a group of $100$ configurations with initial temperature $T_i=4.5$ for $N_l=128$.  We observe that the energy  flows to a fixed point around the equilibrium energy of the system's critical temperature.  This value, indicated by the colored shadow in Fig.~\ref{fig:E-flow-128}, was obtained from the Monte Carlo simulations.  We have checked that for the fixed point all the reconstructed images are identical to the input images pixel by pixel, i.e.\ the reconstruction error is null.
As reference, the energy per spin of saturated configurations is $E_0=-2$, whereas the energy per spin of the system at infinite temperature is $E_\infty=0$. 
The same figure shows another set of configurations (with $T_i=1.5$), that flow to another fixed point, far from the critical region: a perfectly ordered configuration.

\begin{figure}
    \centering
    \includegraphics[width= \linewidth]{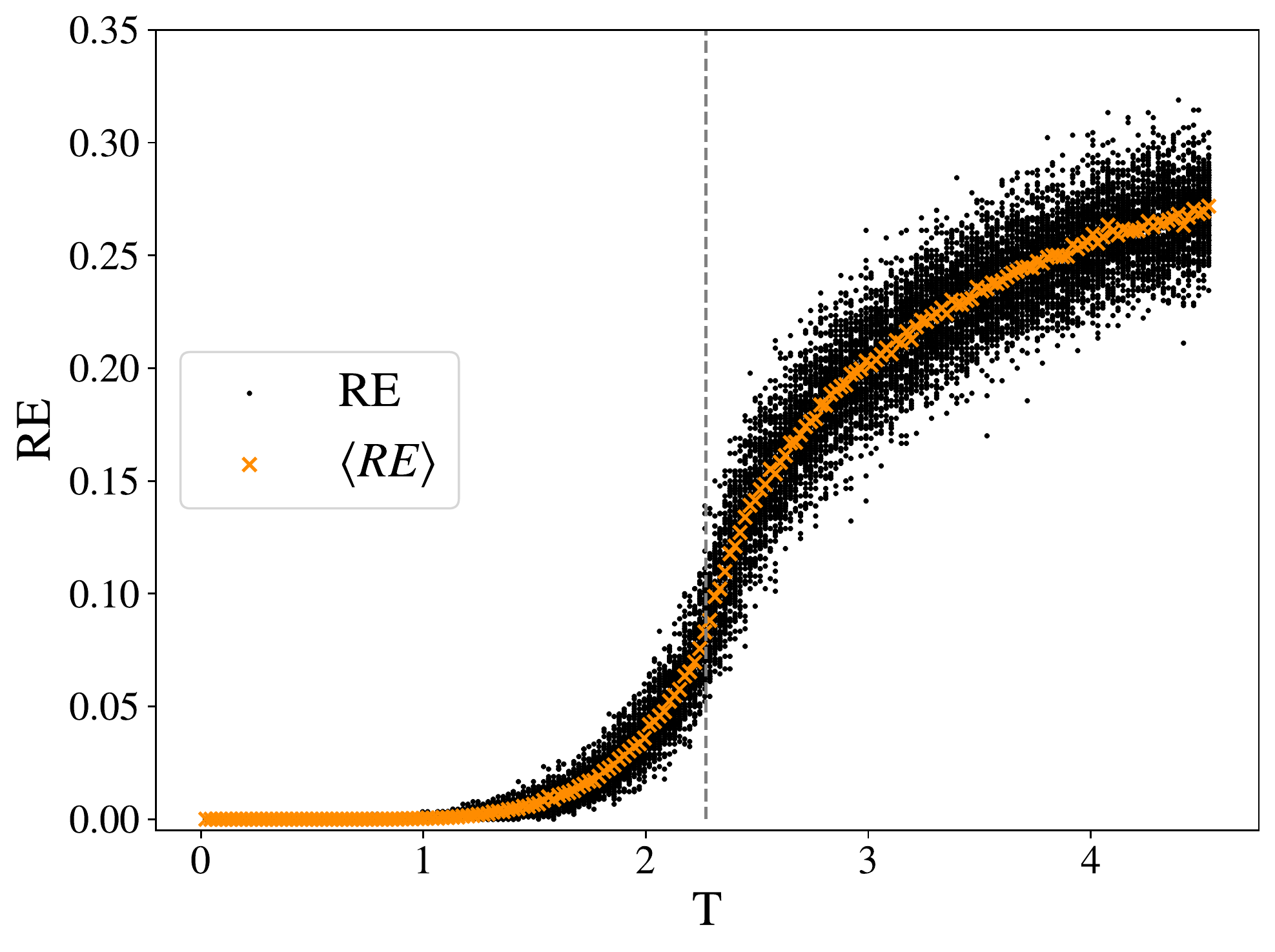}
    \caption{ Reconstruction Error (RE) as a function of temperature. Each black dot is the RE of a single configuration. Orange crosses denote the mean RE at each temperature. The dashed gray vertical line corresponds to the exact critical temperature of the system in the thermodynamic limit.}
    \label{fig:RE-T-128}
\end{figure}

The existence of more than one fixed point can be understood from the curve of RE vs.\ temperature (Fig. \ref{fig:RE-T-128}).  When training a NN, the metrics are typically reported as an average over the whole training and/or validation sets. However, the metrics can display large fluctuations within each set, as it happens in this case: the RE is zero for perfectly ordered snapshots and it increases as the temperature increases.  The critical temperature (vertical dashed line in the figure) roughly coincides with the inflection point in the mean RE curve, as was observed in Ref.~\onlinecite{Acevedo2021}  in the context of anomaly detection.  We must emphasize that the RE in Fig.~\ref{fig:RE-T-128} is \emph{finite} for all configurations that are not fully magnetized ($m=\pm1$): this implies that the fixed point for the high-temperature configurations of Fig.~\ref{fig:E-flow-128}  corresponds to configurations that are physically very unlikely.

\begin{figure}[t]
    \centering
    \includegraphics[width= \linewidth]{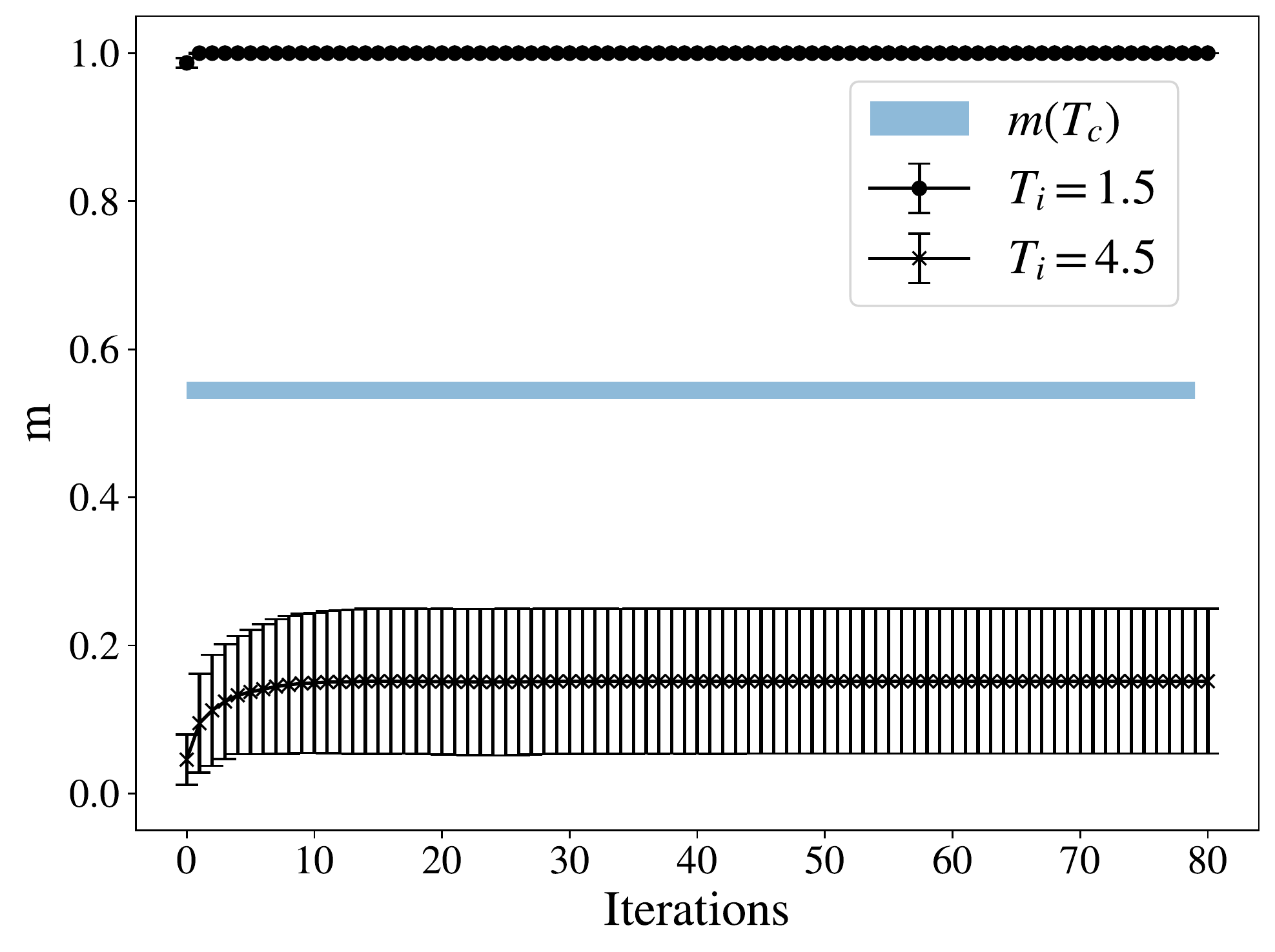}
    \caption{Absolute value of the magnetization flow (per spin) as a function of the number of AE reconstructions, for $100$ independent realizations of the system. Two different initial temperatures $T_i=4.5$ and $T_i=1.5$ are taken to show two different fixed points. The colored shadow denotes the range of magnetizations within dispersion that corresponds to the critical point of the system.}
    \label{fig:m-flow-128-t-cut-0}
\end{figure}

This becomes more clear if we also monitor the magnetization of the configurations along the flow.  Fig.~\ref{fig:m-flow-128-t-cut-0} shows the magnetization flow for $N_l=128$. For $T_i=4.5$ the dispersion is higher than observed for the energy flow.  The fixed point is also observed in the magnetization flow (as it must, since it is actually a fixed point in the flow of configurations),  but the magnetization at the fixed point is quite far from the critical magnetization (recall that at $T_c$, $m=0$ in the thermodynamic limit, but for a finite system $m$ is finite and size-dependent).  This shows that the flux produces configurations that are physically very unlikely, i.e.\ highly improbable when sampling from the Boltzmann distribution at finite temperature.  The fixed-point configurations have an energy close to the critical point energy, but a much lower magnetization.  For $T_i=1.5$, on the other hand, the fixed point is the ground state of the system, where $E=E_0$ and $m=\pm1$. 

\begin{figure}[t]
    \centering
    \includegraphics[width= \linewidth]{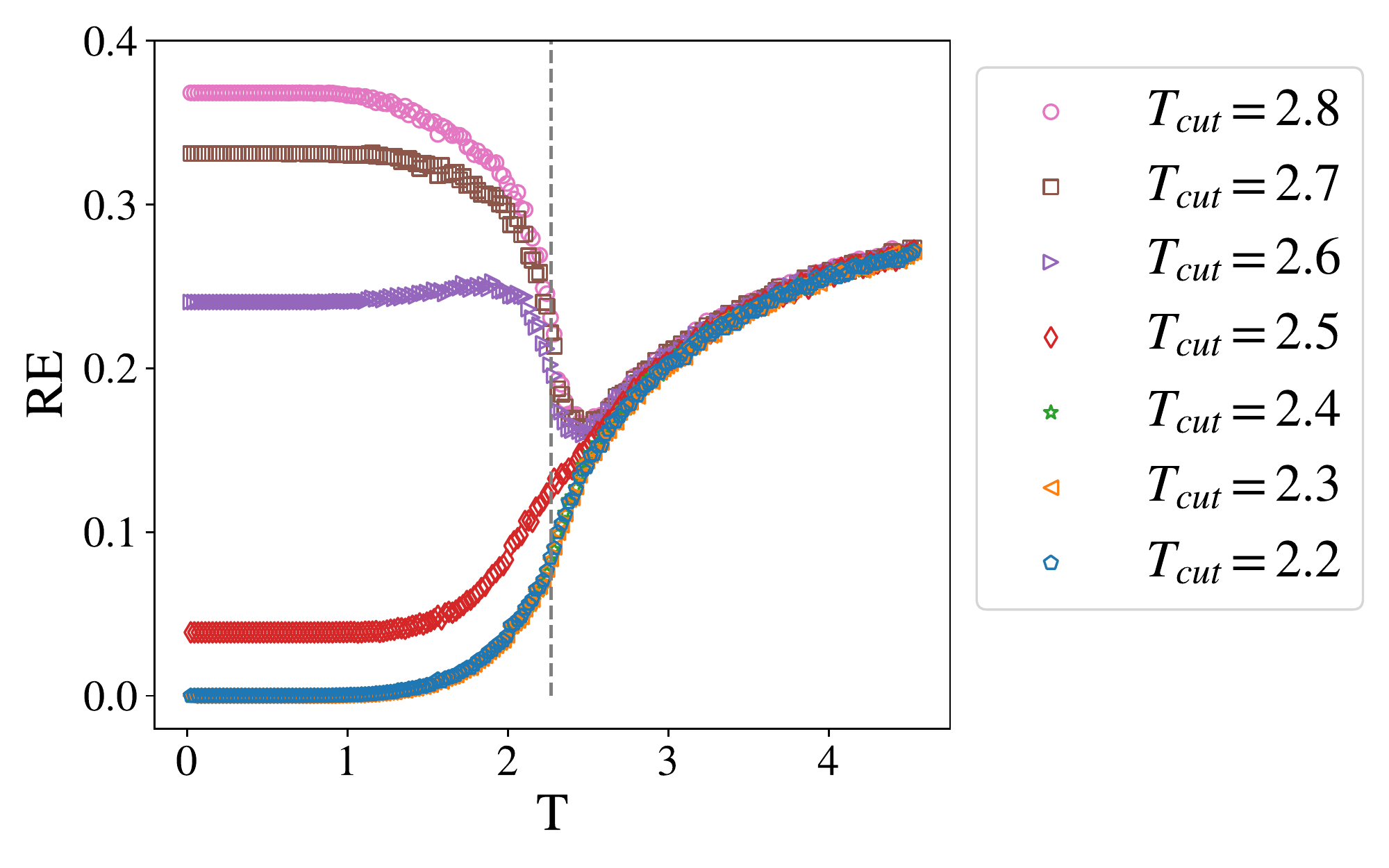}
    \caption{Mean RE as a function of temperature for $N_l=128$ and several values of $T_{cut}$. The dashed gray vertical line corresponds to the exact critical temperature of the system in the thermodynamic limit. The curves with $T_{cut} = 2.2,2.3,2.4$ are overlapped.
    }
    \label{fig:RE_V}
\end{figure}

In Ref.~\onlinecite{giataganas2021neural}, low-temperature configurations were also reported to flow towards the critical point, in contrast to our finding above.  To understand how low-temperature configurations can flow to the critical point, we introduce a parameter $T_{cut}$, which sets a lower bound for the temperatures to be included in the training set.  
This parameter induces strong modifications in the RE as a function of temperature and consequently in the AE flow. 
Fig.~\ref{fig:RE_V} shows that for high $T_{cut}$ the RE rises at low temperatures.
This rise in the RE for ordered configurations is not unexpected, since for this value of $T_{cut}$ the training set includes no ordered configurations. This is similar to the situation encountered using Anomaly Detection\cite{AnomalyDetectionPRL}.   

Interestingly, Fig.~\ref{fig:RE_V} shows that for $T_{cut} \geq 2.6$ the RE is minimum for a temperature that is outside the range of training temperatures, i.e.\ to the left of $T_{cut}$, but in the paramagnetic phase, i.e.\ to the right of $T_c$.  The presence of this minimum is non-trivial and could be related to the fact that the intrinsic dimension ($I_d$) of the input images has a local minimum at $T_c$ \cite{IntrinsicDimension2021}. The $I_d$ is defined in Ref.~\onlinecite{IntrinsicDimension2021} as ``the minimum number of variables needed to accurately describe the important features of a data set''. 
Trivially,  $I_d=1$  at zero temperature, so that a saturated configurations can be perfectly reproduced with only one neuron that learns the order parameter (using an AE), or with only one principal component (using principal component analysis\cite{Wetzel2017}).  At sufficiently high temperature, $I_d=N$, since every spin is independent\cite{IntrinsicDimension2021}. 
Since the AE is doing dimensional reduction in the latent space, it is natural to expect that reconstruction is better when $N_l$ exceeds the intrinsic dimension of data, which is temperature-dependent. 
The final performance of the network will depend both in $N_l$ and training parameters. 

Fig.~\ref{fig:RE_V} is instructive to understand why it is possible to observe the AE-flow head towards configurations that may appear critical. For high temperatures the RE is high because of the high $I_d$ of data. For low temperatures, the RE is also high  because the configurations with small $I_d$ have been removed from the training set.  One would expect that, if the flow has a physical fixed point, it would correspond to a minimum of the curve of Fig.~\ref{fig:RE_V} with an RE very close to 0 (if the fixed point were the critical state, this minimum would be at $T_c$).  Instead, we observe in Fig.~\ref{fig:RE_V} and in general that when the RE curve develops a local minimum, the critical temperature does not match this minimum but roughly coincides with the inflection point in the mean RE curve (as in Fig.~\ref{fig:RE-T-128} and in Ref.~\onlinecite{Acevedo2021}).


\subsection{Energy and magnetization in the reconstructed configurations}

\begin{figure}[t]
    \centering
    \includegraphics[width= \linewidth]{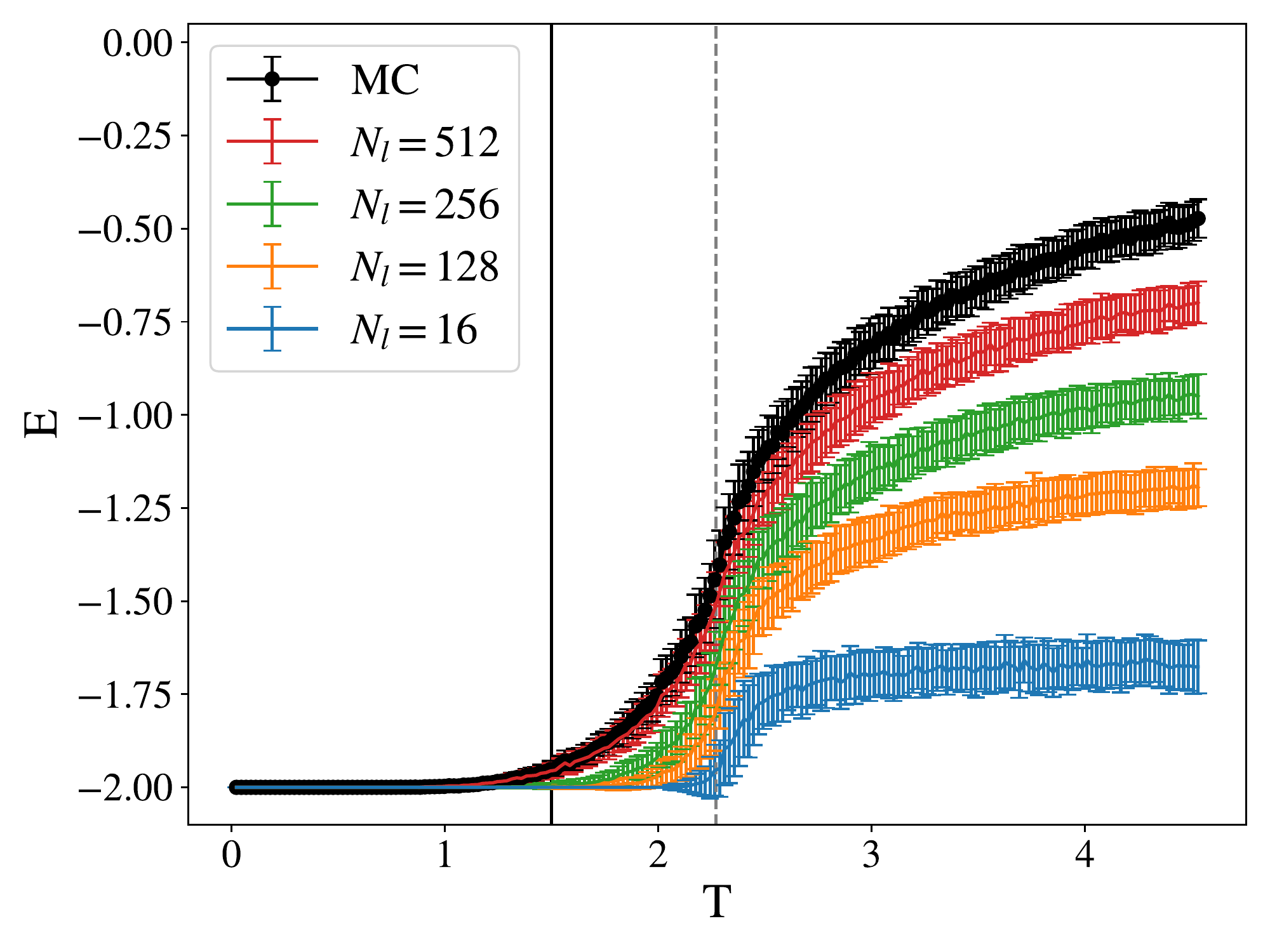}
    \caption{Energy (per spin) as a function of temperature. In black, Monte Carlo sampled configurations. In different colors, the corresponding energies of the first reconstruction made by AEs with different number of neurons in the latent space $(N_l)$. In black solid line, $T_{cut}=1.5$. In gray dashed line the exact critical temperature  of the system in the thermodynamic limit. }
    \label{fig:E_1_T_cut_1.5}
\end{figure}

\begin{figure}[t]
    \centering
    \includegraphics[width= \linewidth]{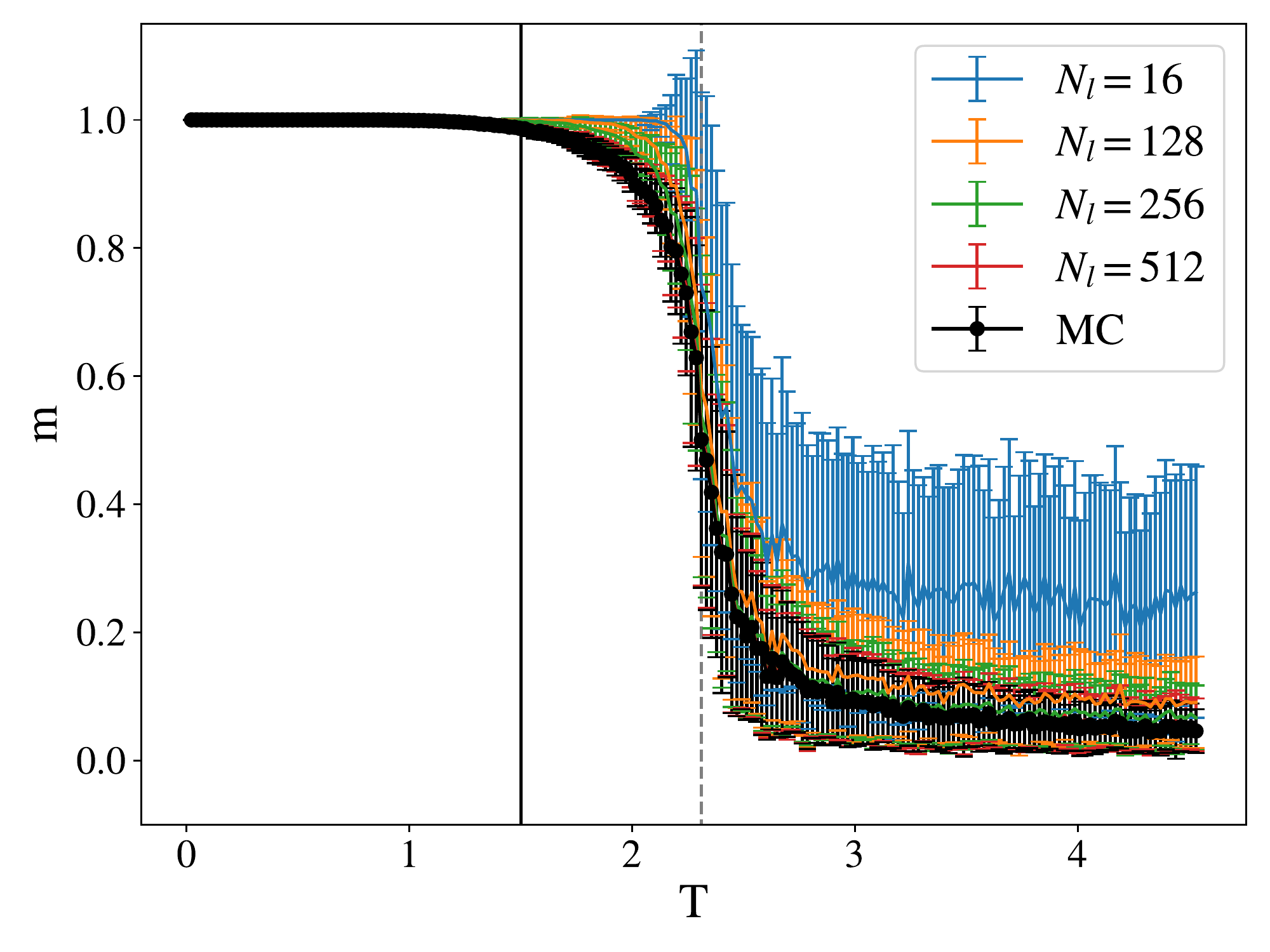}
    \caption{Magnetization per spin as a function of temperature. In black, Monte Carlo sampled configurations. In different colors, the corresponding magnetizations of the first reconstruction made by AEs with different number of neurons in the latent space $(N_l)$. In black solid line, $T_{cut}=1.5$. In gray dashed line the exact critical temperature  of the system in the thermodynamic limit.}
    \label{fig:m_1_T_cut_1.5}
\end{figure}

Fig.~\ref{fig:E_1_T_cut_1.5} shows the MC equilibrium energy vs.\ temperature, together with the energy of the first AE-reconstructed configurations vs.\ the temperature of the initial configuration and for different values of $N_l$.  It is clear that the reconstructed energy is systematically smaller, which agrees with the energy flow from Fig.~\ref{fig:E-flow-128}.  The same plot for the magnetization (Fig.~\ref{fig:m_1_T_cut_1.5}) shows that the reconstructed magnetization is higher than the original for low $N_l$, but can be correctly reproduced with $N_l=512$.  The lower reconstructed energy together with a slightly higher magnetization is consistent with a reconstruction process that does a sort of coarse-graining of the image, blurring interfaces and producing a configuration with larger domains and fewer domain walls, i.e.\ lower energy.

\begin{figure}[H]
    \centering
    \includegraphics[width= \linewidth]{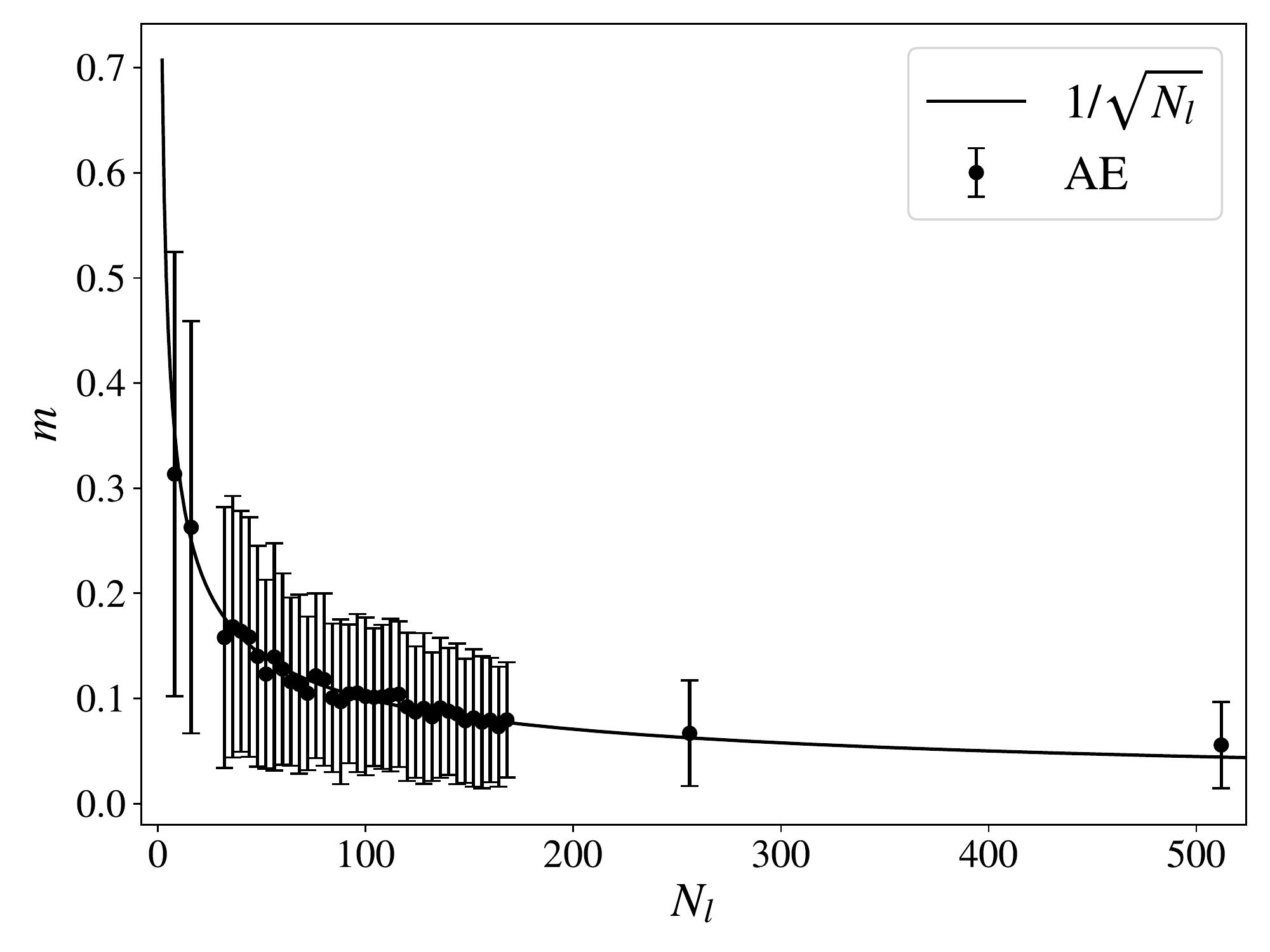}
    \caption{Average reconstructed magnetization for the highest training temperature ($T=4.5$) vs.  $N_l$; i.e., the last value of $m$ in Fig.~\ref{fig:m_1_T_cut_1.5}.  The continuous line is the function $1/\sqrt{N_l}$.}
    \label{fig:sqrtN}
\end{figure}

Fig. \ref{fig:sqrtN} shows the average magnetization at our highest training temperature ($T=4.5$) as a function of $N_l$; i.e., the last value of $m$ in Fig.~\ref{fig:m_1_T_cut_1.5}, for each $N_l$.  For comparison we plot the function $1/ \sqrt{N_l}$, which is the mean absolute value of the magnetization of $N_l$ random spins (and corresponds to the situation of  model \eqref{eq:H} for $T\gg 1$).
We see a good agreement between this two quantities, which could be interpreted as follows. For high enough temperature spins are uncorrelated, and the $N$ random Ising variables have $m\approx 1/\sqrt{N}$ (we have checked that this value agrees within dispersion with the magnetization of our Monte Carlo simulation at our highest training temperature). After a single forward pass through the AE, the number of degrees of freedom is roughly $N_l$ due to the compression in the latent space.  Then, the resulting reconstructed images (of size $N \times N$) are ``as random as possible'' with $N_l$ degrees of freedom, and thus have the same mean magnetization as a colection of only $N_l$ random Ising variables. 

\subsection{Flow dependence with $N_l$ and $T_{cut}$}

\begin{figure}[t]
    \centering
    \includegraphics[width= \linewidth]{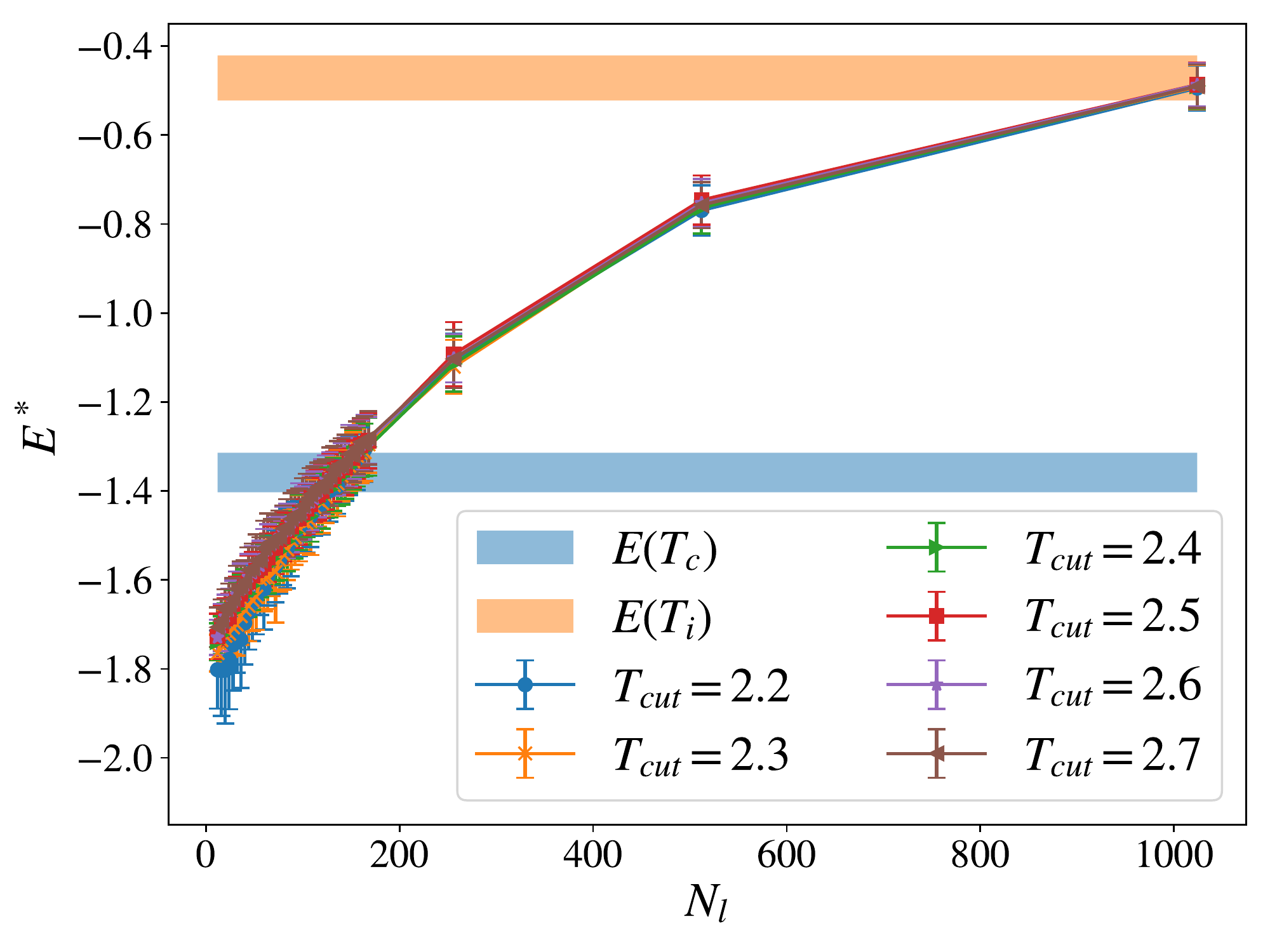}
    \caption{ Final energy of the AE-flow, $E^*$, as a function of the number of variables in the latent space, $N_l$, for several values of $T_{cut}$. The initial group of configurations has a shared temperature of $T_i=4.5$. The blue shadow corresponds to the energy of the system at the critical point within dispersion (1 sigma). The orange shadow corresponds to the energy of the system at $T=T_i$, within dispersion (1 sigma).}
    \label{fig:E_star_h}
\end{figure}

From the above discussion of AE-flow and RE, and given that for sufficiently high $N_l$ the reconstructions should be good at all temperatures, one would expect that for high enough $N_l$ every configuration should be a fixed point, or very close to one.  Fig.~\ref{fig:E_star_h} shows how the fixed-point energy $E^*$ (the energy at the end of the flow) depends on both $N_l$ and $T_{cut}$ for configurations with initial temperature $T_i=4.5$.  $E^*$ increases monotonically with the number neurons in the latent space.  We see thus that configurations in general do not flow towards the critical region, but that one can find a network and a training scheme that stops both below or above the critical energy range.

The behavior of $E^*$ with $N_l$ for flows starting with configurations at $T_i=1.5$ is shown in Fig.~\ref{fig:E_star}.  We see again that for high enough $N_l$ the initial configurations are fixed points, and that  $E^*$ is strongly dependent on both $N_l$ and $T_{cut}$.

Finally, Fig.~\ref{fig:RE_star} presents the mean RE for perfectly ordered snapshots and different values of $N_l$ and $T_{cut}$.  This shows two regimes of the AE, where it can or cannot reproduce ordered configurations, and the crossover between them.  Even if the ordered phase is excluded from training, the network can learn how to reconstruct ordered configurations if $N_l$ is high enough.
This result may be interpreted in at least two related ways. On one hand it must be taken into account that even when training in the paramagnetic phase, the system presents short-range ferromagnetic order. This means that the network may have seen and learned to reproduce some locally-ordered domains. On the other hand, as the network has more and more parameters and it is trained to reproduce disordered snapshots, it is trying to achieve a very difficult task. Then, it may be natural for it to also have learned to reproduced perfectly ordered snapshots, which is a much simpler task. A similar situation appears in the anomaly detection context in Ref. \cite{Acevedo2021} (Figs. 9 and 10).

\section{Conclusions}

We have studied how spin configurations of the ferromagnetic 2-$d$ Ising model transform under the flow generated by successive image reconstructions done with a fully connected AE (AE-flow\cite{giataganas2021neural}).  To monitor the flow we chose to follow the order parameter and energy along the successive transformations.  We found that these two quantities behave differently, and depend strongly on the network hyperparameters.  Using an architecture with a single hidden layer with $N_l$ neurons, we found a value of $N_l$ where the final energy of the flow coincides within dispersion with the range of energies that corresponds to the critical temperature of the system (Fig. \ref{fig:E-flow-128}). Nonetheless, for the same configurations the value of the order parameter (the magnetization per spin, $m$) did not agree with the expected physical value (Fig.~\ref{fig:m-flow-128-t-cut-0}), which is zero in the thermodynamic limit but finite for a finite number of spins.  This shows that the configurations produced by the flow are not physical, i.e.\ are very unlikely to be found when sampling from the Boltzmann distribution. 

In Ref.~\onlinecite{giataganas2021neural} the flow was monitored with a separate neural network, trained to assign a temperature to each spin configuration.  These authors found that under certain conditions the temperatures in the AE-flow, as assigned by the second neural network, converge to the critical temperature of the system, $T_c$.  Although this neural network approach is interesting, it does not actually allow to check whether the transformed configurations are near the critical point. 
Nonetheless, if one insists in monitoring the flow using neural networks, the magnetization, energy and correlation length can be estimated using neural network as regression models\cite{civitciouglu2021machine}.

\begin{figure}[t]
    \centering
    \includegraphics[width= \linewidth]{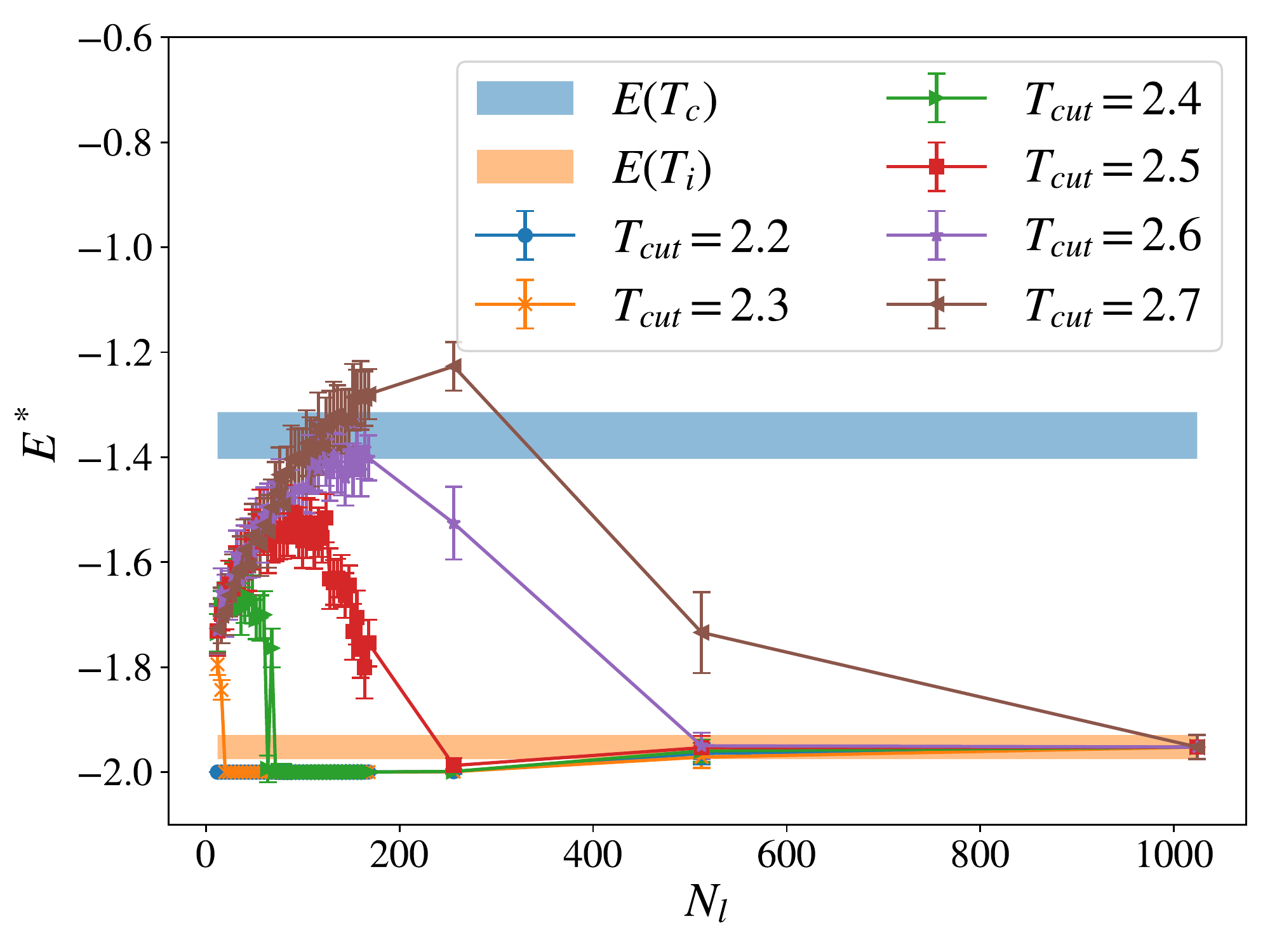}
    \caption{ Final energy of the AE-flow, $E^*$, as a function of the number of variables in the latent space, $N_l$, for several values of $T_{cut}$. The initial group of configurations has a shared temperature of $T_i=1.5$. The blue shadow corresponds to the energy of the system at the critical point within dispersion (1 sigma). The orange shadow corresponds to the energy of the system at $T=T_i$, within dispersion (1 sigma). }
    \label{fig:E_star}
\end{figure}

\begin{figure}[t!]
    \centering
    \includegraphics[width= \linewidth]{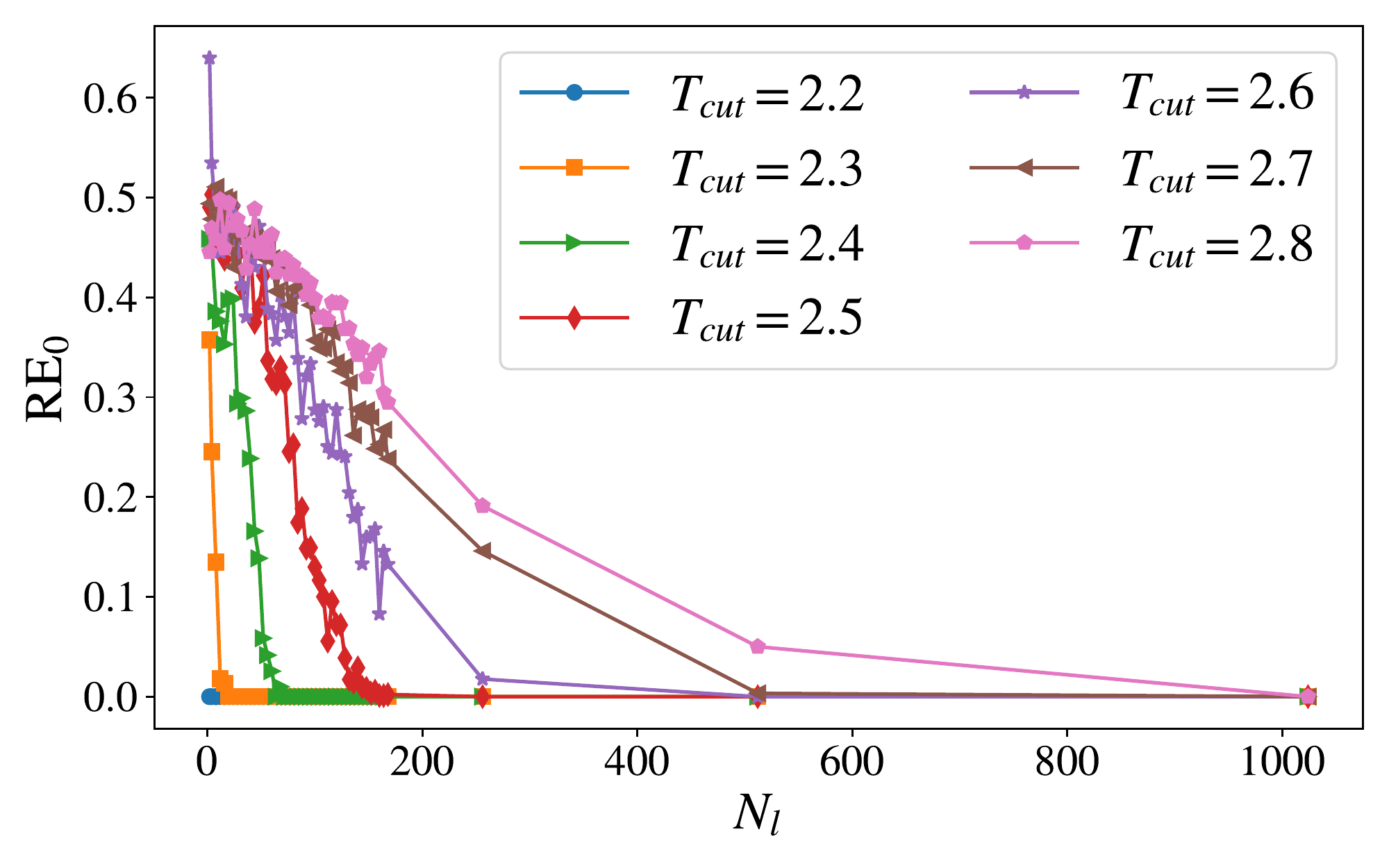}
    \caption{Mean RE computed on the reconstruction of perfectly ordered configurations (here defined as $\text{RE}_0$), for different values of $N_l$ and $T_{cut}$.}
    \label{fig:RE_star}
\end{figure}

Training an AE with temperatures in the range $[0.02,4.5]$ we found that the quality of the reconstructions is very inhomogeneous: 
the reconstruction error (RE) is strongly temperature-dependent, being zero for saturated configurations ($m=\pm1$), and having an inflection point 
around $T_c\approx 2.269$ (the same was found in an Anomaly Detection study using AEs in similar spin systems\cite{Acevedo2021}).  
As a consequence, configurations from the ferromagnetic phase flow to the saturated state.  In order to study non-saturated fixed points in the AE-flow, we introduced a temperature $T_{cut}$, a lower bound for the temperature of the data included in training, and we found that $T_{cut}$ allows for the formation of a non-trivial local minima of RE as a function of temperature near the critical region (Fig. \ref{fig:RE_V}).  We argue that this minimum should be related to the minimum in the intrinsic dimension (see Sec.~\ref{sec:Results-AE-flow}) of data in $T_c$\cite{IntrinsicDimension2021}. Finally, we showed that there is a strong dependence of the flow on both $T_{cut}$ and  $N_l$ (Figs.~\ref{fig:E_star_h} and~\ref{fig:E_star}).  These results contradict the claim\cite{giataganas2021neural} that the configurations flow towards the critical region: the energy fixed point can be tuned to be higher or lower than the critical energy, depending on the choice $N_l$ and $T_{cut}$.  In any case, even if the hyperparameters are tuned to drive the energy to the critical range, the configurations are not critical in the sense that they are not representative of the Boltzmann ensemble at the critical temperature, as mentioned above.

We interpret the NN-flow in terms of the reconstruction error (RE), since the modification of spin configurations implies necessarily a finite RE, and in our case a fixed point in the flow also implies a vanishing RE.  An interesting and more general situation could be where the fixed point is statistical, i.e., the physical quantities are fixed within dispersion, but the RE is finite.  Whether this situation can be realized with some neural network remains to be established. To realize this scenario, it could be fruitful to implement a set of metrics with physical content, like a cost function that penalizes the error in energy, order parameter, and/or correlation length.  This is a possible line to pursue in future studies.  We hope that our work motivates further research on the neural network flow of spin configurations.



\section{Data availability}

The raw data required to reproduce these findings are available to download from \href{http://dx.doi.org/10.17632/53bshnvwp4.1}{http://dx.doi.org/10.17632/53bshnvwp4.1}. The codes used in this work are available to download from  \href{https://github.com/acevedo-s/NN-flow}{https://github.com/acevedo-s/NN-flow}

\bibliographystyle{unsrt}
\bibliography{references.bib} 

\begin{thebibliography}{10}

\bibitem{carrasquilla2017machine}
Juan Carrasquilla and Roger~G Melko.
\newblock Machine learning phases of matter.
\newblock {\em Nature Physics}, 13(5):431--434, 2017.

\bibitem{van2017learning}
Evert~PL Van~Nieuwenburg, Ye-Hua Liu, and Sebastian~D Huber.
\newblock Learning phase transitions by confusion.
\newblock {\em Nature Physics}, 13(5):435--439, 2017.

\bibitem{CORTE2021}
I.~Corte, S.~Acevedo, M.~Arlego, and C.A. Lamas.
\newblock Exploring neural network training strategies to determine phase
  transitions in frustrated magnetic models.
\newblock {\em Computational Materials Science}, 198:110702, 2021.

\bibitem{Ponte2017}
Pedro Ponte and Roger~G. Melko.
\newblock Kernel methods for interpretable machine learning of order
  parameters.
\newblock {\em Phys. Rev. B}, 96:205146, Nov 2017.

\bibitem{Wang2016}
Lei Wang.
\newblock Discovering phase transitions with unsupervised learning.
\newblock {\em Phys. Rev. B}, 94:195105, Nov 2016.

\bibitem{Wang2017}
Ce~Wang and Hui Zhai.
\newblock Machine learning of frustrated classical spin models. i. principal
  component analysis.
\newblock {\em Phys. Rev. B}, 96:144432, Oct 2017.

\bibitem{IntrinsicDimension2021}
T.~Mendes-Santos, X.~Turkeshi, M.~Dalmonte, and Alex Rodriguez.
\newblock Unsupervised learning universal critical behavior via the intrinsic
  dimension.
\newblock {\em Phys. Rev. X}, 11:011040, Feb 2021.

\bibitem{fischer2006predicting}
Christopher~C Fischer, Kevin~J Tibbetts, Dane Morgan, and Gerbrand Ceder.
\newblock Predicting crystal structure by merging data mining with quantum
  mechanics.
\newblock {\em Nature materials}, 5(8):641--646, 2006.

\bibitem{Crystal2003}
Stefano Curtarolo, Dane Morgan, Kristin Persson, John Rodgers, and Gerbrand
  Ceder.
\newblock Predicting crystal structures with data mining of quantum
  calculations.
\newblock {\em Phys. Rev. Lett.}, 91:135503, Sep 2003.

\bibitem{samarakoon2020}
Anjana~M. Samarakoon, Kipton Barros, Ying~Wai Li, Markus Eisenbach, Qiang
  Zhang, Feng Ye, V.~Sharma, Z.~L. Dun, Haidong Zhou, Santiago~A. Grigera,
  Cristian~D. Batista, and D.~Alan Tennant.
\newblock Machine-learning-assisted insight into spin ice {{Dy2Ti2O7}}.
\newblock {\em Nat Commun}, 11(1):892, February 2020.

\bibitem{RBM-MonteCarlo}
Li~Huang and Lei Wang.
\newblock Accelerated monte carlo simulations with restricted boltzmann
  machines.
\newblock {\em Phys. Rev. B}, 95:035105, Jan 2017.

\bibitem{koch2018mutual}
Maciej Koch-Janusz and Zohar Ringel.
\newblock Mutual information, neural networks and the renormalization group.
\newblock {\em Nature Physics}, 14(6):578--582, 2018.

\bibitem{RG-ML}
Shuo-Hui Li and Lei Wang.
\newblock Neural network renormalization group.
\newblock {\em Phys. Rev. Lett.}, 121:260601, Dec 2018.

\bibitem{torlai2018neural}
Giacomo Torlai, Guglielmo Mazzola, Juan Carrasquilla, Matthias Troyer, Roger
  Melko, and Giuseppe Carleo.
\newblock Neural-network quantum state tomography.
\newblock {\em Nature Physics}, 14(5):447--450, 2018.

\bibitem{carleo2017}
Giuseppe Carleo and Matthias Troyer.
\newblock Solving the quantum many-body problem with artificial neural
  networks.
\newblock {\em Science}, 355(6325):602--606, 2017.

\bibitem{Review-ML-physics}
Giuseppe Carleo, Ignacio Cirac, Kyle Cranmer, Laurent Daudet, Maria Schuld,
  Naftali Tishby, Leslie Vogt-Maranto, and Lenka Zdeborov\'a.
\newblock Machine learning and the physical sciences.
\newblock {\em Rev. Mod. Phys.}, 91:045002, Dec 2019.

\bibitem{RBMFlow0}
Satoshi Iso, Shotaro Shiba, and Sumito Yokoo.
\newblock Scale-invariant feature extraction of neural network and
  renormalization group flow.
\newblock {\em Phys. Rev. E}, 97:053304, May 2018.

\bibitem{RBMFlow}
Shotaro~Shiba Funai and Dimitrios Giataganas.
\newblock Thermodynamics and feature extraction by machine learning.
\newblock {\em Phys. Rev. Research}, 2:033415, Sep 2020.

\bibitem{giataganas2021neural}
Dimitrios Giataganas, Ching-Yu Huang, and Feng-Li Lin.
\newblock Neural network flows of low q-state potts and clock models, 2021.

\bibitem{newman1999monte}
M~Newman and G~Barkema.
\newblock {\em Monte carlo methods in statistical physics chapter 1-4},
  volume~24.
\newblock Oxford University Press: New York, USA, 1999.

\bibitem{Goodfellow-et-al-2016}
Ian Goodfellow, Yoshua Bengio, and Aaron Courville.
\newblock {\em Deep Learning}.
\newblock MIT Press, 2016.
\newblock \url{http://www.deeplearningbook.org}.

\bibitem{chollet2021deep}
Francois Chollet.
\newblock {\em Deep learning with Python}.
\newblock Simon and Schuster, 2021.

\bibitem{tensorflow_developers_2021_5799851}
TensorFlow Developers.
\newblock Tensorflow, dec 2021.

\bibitem{Acevedo2021}
S.~Acevedo, M.~Arlego, and C.~A. Lamas.
\newblock Phase diagram study of a two-dimensional frustrated antiferromagnet
  via unsupervised machine learning.
\newblock {\em Phys. Rev. B}, 103:134422, Apr 2021.

\bibitem{AnomalyDetectionPRL}
Korbinian Kottmann, Patrick Huembeli, Maciej Lewenstein, and Antonio Ac\'{\i}n.
\newblock Unsupervised phase discovery with deep anomaly detection.
\newblock {\em Phys. Rev. Lett.}, 125:170603, Oct 2020.

\bibitem{Wetzel2017}
Sebastian~J. Wetzel.
\newblock Unsupervised learning of phase transitions: From principal component
  analysis to variational autoencoders.
\newblock {\em Phys. Rev. E}, 96:022140, Aug 2017.

\bibitem{civitciouglu2021machine}
Burak {\c{C}}ivitcio{\u{g}}lu, Rudolf~A R{\"o}mer, and Andreas Honecker.
\newblock Machine learning the square-lattice ising model.
\newblock {\em arXiv preprint arXiv:2111.13413}, 2021.

\end{thebibliography}

\end{document}